\documentclass[prx,twocolumn,showpacs,groupedaddress,longbibliography,10pt]{revtex4-2} 

\usepackage{lipsum, babel}

\usepackage{siunitx}
\usepackage{amsmath,bm}
\usepackage{amsbsy}
\usepackage{amssymb}
\usepackage{mathptmx, textcomp}
\usepackage{color}
\usepackage{braket}
\usepackage{graphicx}
\usepackage{textcomp}
\usepackage{notes2bib}
\usepackage{textcomp}
\usepackage{mwe}
\usepackage{amsmath,amssymb,graphicx}
\usepackage{color,times}
\definecolor{darkblue}{rgb}{0, 0, 0.8}
\usepackage[colorlinks=true, breaklinks=true, linkcolor=darkblue, citecolor=darkblue, urlcolor=darkblue]{hyperref}
\graphicspath{{..//figures////final_figures//}}
\DeclareMathAlphabet{\mathcal}{OMS}{cmsy}{m}{n}

\begin{document}

\title{Microwave-engineering of programmable XXZ Hamiltonians in arrays of Rydberg atoms}

\author{P. Scholl}
\altaffiliation{These authors contributed equally to this work.}
\author{H. J. Williams$^*$}
\author{G. Bornet$^*$}
\author{F. Wallner}
\altaffiliation{Also at Department of Physics, Technical University of Munich, James-Franck-Strasse 1, 85748 Garching, Germany}
\author{D. Barredo}
\altaffiliation{Also at Nanomaterials and Nanotechnology Research Center (CINN‑CSIC), 
Universidad de Oviedo (UO), Principado de Asturias, 33940 El Entrego, Spain}
\author{T.~Lahaye}
\author{A.~Browaeys}
\email{antoine.browaeys@institutoptique.fr}
\affiliation{Universit\'e Paris-Saclay, Institut d'Optique Graduate School,\\
CNRS, Laboratoire Charles Fabry, 91127 Palaiseau Cedex, France}
\author{L.~Henriet}
\author{A.~Signoles}
\affiliation{Pasqal, 2 avenue Augustin Fresnel, 91120 Palaiseau, France}
\author{C.~Hainaut}
\author{T.~Franz}
\author{S.~Geier}
\author{A.~Tebben}
\author{A.~Salzinger}
\author{G.~Z\"urn}
\author{M.~Weidem\"uller}
\affiliation{Physikalisches Institut, Universit\"at Heidelberg, Im Neuenheimer Feld 226, 69120 Heidelberg, Germany}

\date{\today}

\begin{abstract}

We use the resonant dipole-dipole interaction between Rydberg atoms and a periodic external microwave 
field to engineer XXZ spin Hamiltonians with tunable anisotropies. 
The atoms are placed in 1D and 2D arrays of optical tweezers. As illustrations, we apply this engineering to
two iconic situations in spin physics: the Heisenberg
model in square arrays and spin transport in 1D. 
We first benchmark the Hamiltonian engineering for two atoms, and then demonstrate the freezing of the magnetization 
on an initially magnetized 2D array. Finally, we explore the dynamics of 1D domain 
wall systems with both periodic and open boundary conditions. 
We systematically compare our data with numerical simulations and assess the residual limitations 
of the technique as well as routes for improvements. The geometrical versatility of the platform, 
combined with the flexibility of the simulated Hamiltonians, opens exciting prospects in the field of quantum 
simulation, quantum information processing and quantum sensing. 

\end{abstract}

\maketitle

\section{Introduction}

Quantum simulation using synthetic quantum systems is now becoming a fruitful approach to 
explore open questions in many-body physics \cite{RMP_Qsim}. Experimental platforms that have been
used for quantum simulation so far include ions~\cite{Blatt2012,MonroeRMP2021}, 
molecules~\cite{Zhou2011,Yan2013}, atoms~\cite{Bloch2012,Gross17} 
or quantum circuits~\cite{Houck2012,RevQcircuit2020}. These systems naturally implement particular instances of  
many-body Hamiltonians, such as the ones describing the interactions between spins or the 
Bose- and Fermi-Hubbard Hamiltonians \cite{Gross17}. Each platform already features 
a high degree of programmability, with the possibility to tune many of the parameters of the simulated Hamiltonians.
In the quest for fully programmable quantum simulators, 
one would like to extend the capabilities to simulate Hamiltonians beyond the ones naturally implemented. 
In this spirit, applying a periodic drive to a system allows for the engineering of a broader class of 
Hamiltonians, where additional parameters can be modified at will. 
This Floquet engineering technique~\cite{Goldman2014}, initially introduced in the context of 
NMR~\cite{Shirley1965,Vandersypen2005}, has been used for digital quantum 
simulation~\cite{Salathe2015} and to explore new 
physical phenomena such as dynamical phase transitions~\cite{Jurcevic2017}, Floquet-
prethermalization~\cite{Peng2021,Rubio2020}, novel phases of matter \cite{Kyprianidis2021} 
and topological configurations~\cite{Aidelsburger2013,Flaschner2016,Meinert2016,
Schweizer2019,Eckardt2017,Wintersperger2020}.

Among the platforms being developed, the one based on Rydberg atoms held in arrays of 
optical tweezers is  a promising candidate for quantum simulation \cite{Browaeys2020} and 
computation \cite{Henriet2020,Morgado2021}. Recent works have demonstrated its potential 
through the implementation of different spin models. 
Firstly, an ensemble of Rydberg atoms coupled by the van der Waals interaction naturally 
realizes the quantum transverse field Ising model. 
Using this fact, arrays containing up to
hundreds of atoms have  been used to prepare antiferromagnetic order in 2D 
\cite{Bakr2018,Lienhard2018,Scholl2021} or 3D~\cite{Song2021}, 
study exotic phases  and quantum phase transitions \cite{Keesling2019,Ebadi2021}, 
and observe the first evidence of a spin liquid~\cite{Semeghini2021}. 
Secondly, the resonant dipole-dipole interaction between Rydberg atoms in states with opposite parity 
implements an XX spin Hamiltonian, which has been used to realize a 
density-dependent Peierls phase \cite{Lienhard2020} and to prepare a symmetry-protected 
topological phase in 1D \cite{deLeseleuc2019}. 
Finally, the dipolar interaction for two Rydberg atoms in states with the same parity leads 
to a XXZ spin Hamiltonian with anisotropy {\it fixed} by the choice of the principal quantum number 
\cite{Whitlock2017}, as demonstrated  in a gas of cold atoms~\cite{Signoles2021}. 
Circular Rydberg atoms also offer the promise of realizing the XXZ model with anisotropy tunable
by external electric and magnetic fields~\cite{Nguyen2018}.

Besides these naturally implemented models, more general spin models, such as XYZ models, 
which can feature either $\rm{SU}(2)$, $\rm{U}(1)$ or even absence of unitary symmetries, are also of 
general interest to study ground-state~\cite{Dmitriev2002} 
and out-of-equilibrium many-body physics~\cite{Wei2018}. 
In this context, transport properties of spin excitations are actively 
studied, both experimentally and theoretically (e.g.~\cite{Cheneau2012,Gobert2005,Sirker2009,Barmettler2009,Bertini2021}). 
For 1D systems, the behavior is known to be highly dependent on the parameters of the 
Hamiltonians~\cite{Giamarchi1D}. Several experimental methods, involving the relaxation of 
spin-spiral states~\cite{Jepsen2020,Hild2014} or the melting of initially prepared 
domain walls~\cite{wei2021,Joshi2021}, enable the extraction of global transport behaviors ranging from ballistic 
to localized ones as a function of the Hamiltonian parameters. 
Furthermore, the experimental development of single-atom resolution techniques gives access
to the exploration of transport properties through correlation functions, 
as demonstrated with trapped ions ~\cite{Richerme2014,Jurcevic2014,Tan2021}, or ultra-cold atoms
in optical lattices~\cite{Fukuhara2013}. 

Programmable XXZ Hamiltonians have been recently demonstrated on a periodically driven 
Rydberg gas where the atoms are coupled by the resonant dipole-dipole interaction~\cite{Geier2021}. 
This technique offers the opportunity to arbitrarily and dynamically tune the anisotropy of the applied Hamiltonian.
However, the use of a gas in~\cite{Geier2021} prevented the direct observation of the underlying coherent dynamics. 
Here, we extend this demonstration to the case of {\it ordered} arrays of Rydberg atoms with individual 
addressing and measurement capabilities.
The versatility and control of the platform allows us to implement the XXZ Hamiltonian 
in several situations, ranging from 1D with open or periodic boundary conditions to 2D geometries. 
This enables us to explore {\it coherent} spin transport in a few-body system through 
the investigation of domain wall melting experiments.

\section{Microwave engineering of XXZ Hamiltonians}\label{Sec:Theo_engineer} 

In this first section, we apply the average Hamiltonian theory to the specific case of Rydberg atoms and briefly show 
how to engineer the XXZ spin model with tunable parameters. 
We closely follow  the approach developed in Ref.~\cite{Vandersypen2005,Geier2021}. 

We consider an array of Rydberg atoms, each described as a two-level system 
with states of opposite parity mapped onto pseudo-spin states: 
$\ket{nS} = \ket{\downarrow}$ and $\ket{nP} = \ket{\uparrow}$. 
The resonant dipole-dipole interaction couples the atoms, leading to the XX Hamiltonian:
\begin{equation}
H_{\rm XX}={1\over 2}\sum_{i\neq j} J_{ij} (\sigma^x_i \sigma^x_j + \sigma^y_i \sigma^y_j).
\end{equation}
Here, $J_{ij}=C_3(1-3\cos^2\theta_{ij})/(2r_{ij}^3)$, where $r_{ij}$ is the distance between atoms $i$ and $j$,  
$\theta_{ij}$ gives their angle compared to the quantization axis, 
and $\sigma^x_i=\ket{\uparrow}\bra{\downarrow}_i+\ket{\downarrow}\bra{\uparrow}_i$ 
and $\sigma^y_i=i(\ket{\uparrow}\bra{\downarrow}_i-\ket{\downarrow}\bra{\uparrow}_i)$ 
are the Pauli matrices for atom $i$. 
Adding a resonant microwave field to couple the $\ket{\downarrow}$ and $\ket{\uparrow}$ states, 
the Hamiltonian becomes, in the rotating-wave approximation:
\begin{equation}\label{Eq:Hdriven}
H_{\text{driven}}= H_{\rm XX} + \dfrac{\hbar\Omega(t)}{2}\sum_i \cos\phi(t)\, \sigma^x_i + \sin\phi(t)\, \sigma^y_i,
\end{equation}
where $\Omega(t)$ and $\phi(t)$ are the Rabi frequency and phase of the microwave field, respectively.
We use a sequence $(X, -Y, Y, -X)$ of four  $\pi /2$-Gaussian pulses with constant phases $\phi = (0, -\pi /2 , \pi /2, \pi)$ 
separated by durations $\tau_{1,2}$ and $2\tau_3$, shown in Fig.~\ref{fig:twoatoms}(a). 
The time-average of $H_{\text{driven}}$ over a sequence leads to the time-independent Hamiltonian $H_{\rm av}$:
\begin{equation}\label{Eq:HFloquet}
\begin{aligned}
H_{\rm av}={1\over 2}\sum_{i\neq j}\dfrac{2J_{ij}}{t_c} [ (\tau_1 + \tau_2)\,\sigma^x_i \sigma^x_j + 
(\tau_1 + \tau_3)\,\sigma^y_i \sigma^y_j\\ + (\tau_2 + \tau_3)\, \sigma^z_i \sigma^z_j ]\  ,
\end{aligned}
\end{equation}
where $t_c = 2(\tau_1 + \tau_2 + \tau_3)$ is the total duration of the sequence. 
The dynamics of the system is governed in good approximation by $H_{\rm av}$ 
when the duration of each pulse is negligible with respect to $t_c$. Moreover, 
$t_c$ needs to be much shorter than the interaction timescales set by the 
averaged interaction energy $J_m = 1/N \sum_{i\neq j} J_{ij}$, with $N$ the total number of spins. This leads to 
the requirement $J_m  t_c \ll 2 \pi$. 
As the number of nearest neighbours, and hence $J_m$, depends on the geometry of the array,
$t_c$ must be adapted accordingly.
Equation~(\ref{Eq:HFloquet}) has the form of an XYZ Hamiltonian, whose coefficients are tunable 
by simply varying the delays between the pulses. 
In this work we restrict ourselves to the case of the XXZ Hamiltonian, which conserves
the number of spin excitations:
\begin{equation}\label{Eq:HXXZ}
H_{\rm XXZ}={1\over 2}\sum_{i\neq j} J_{ij}^x(\sigma^x_i \sigma^x_j + \sigma^y_i \sigma^y_j) + J_{ij}^z\sigma^z_i \sigma^z_j\ , 
\end{equation}
where $J_{ij}^x = J_{ij}^y =2 J_{ij}(\tau_1 + \tau_2)/t_c$ and $J_{ij}^z = 4J_{ij}\tau_2/t_c$, with $\tau_2 = \tau_3$.
The anisotropy of the Hamiltonian $\delta = J_{ij}^z/J_{ij}^x = 2 \tau_2 /(\tau_1 + \tau_2)$ is thus tunable 
in the range $0 < \delta < 2$. The nearest-neighbor interaction energies $J_x, J_z$ in the engineered XXZ 
model are related to the nearest-neighbor
interaction energy $J$ by: $J_x(\delta)=2J/(2+\delta)$ and $J_z(\delta)=2J\, \delta/(2+\delta)$.

\section{Experimental setup and procedures}\label{Sec:Expsetup}

Our experimental setup 
is based on arrays of single $^{87}\text{Rb}$ atoms trapped in optical tweezers \cite{Barredo2018,Nogrette2014,Schymik2020}. 
The atoms are initialized in their ground state $\ket{g} = \ket{5S_{1/2}, F = 2, m_F = 2}$ 
by optical pumping (efficiency $\sim99.5 \%$). 
We then switch off the tweezers, and transfer the atoms into the 
$\ket{\downarrow} = \ket{nS_{1/2}, m_J = 1/2}$ Rydberg state using a 
STImulated Raman Adiabatic Passage~\cite{deLeseleuc2019}
involving two lasers tuned on the $5S_{1/2}-6P_{3/2}$ transition at 421~nm 
and $6P_{3/2}-nS_{1/2}$ transition at 1013~nm, respectively (efficiency $\sim 95\%$). 

The microwave field couples the state $\ket{\downarrow}$
to a chosen Zeeman state $\ket{\uparrow}$ of the  $nP_{3/2}$ manifold, in the presence of a 25-G magnetic field. 
This field is parallel to the interatomic axis for the two-atom situation, and 
perpendicular to the plane of the atomic arrays for the remaining experiments, to ensure isotropic interactions.  
The microwave field at a frequency $\omega_{\rm MW}/(2\pi)$ ranging from $5-10$~GHz
is obtained by mixing a microwave signal generated by a synthesizer with the field produced by an 
Arbitrary Waveform Generator \footnote{Tabor Electronics Ltd. SE5081} operating near 200~MHz.
 
To initialise the system in a chosen spin state we address specific
sites within the array~\cite{Deleseleuc2017}. For this purpose, we use a Spatial Light Modulator 
which imprints a specific phase pattern on a 1013~nm laser beam tuned on resonance with the $6P-nS$
transition. This results in a set of focused laser beams (waist $\sim 2\, \mu$m) in the atomic plane, 
whose geometry corresponds to the subset of sites we wish to address, preventing the addressed 
atoms from interacting with the microwaves thanks to the Autler-Townes splitting of the $nS $ state.
We combine this addressing technique with resonant microwave rotations to excite the
targeted atoms to the state $\ket{\uparrow}$, with the others in $\ket{\downarrow}$. 
The fidelity of this preparation  is $\sim 95\%$ per atom.

Following the implementation of a particular sequence, we read out the state of the atoms. 
To do so, we  use the 1013~nm STIRAP laser to de-excite the atoms in the $nS_{1/2}$ state to the $6P_{3/2}$
state from which they decay back to the ground states, and are recaptured in their 
tweezer~\footnote{For the experiments beyond two atoms, prior to sending the 1013 nm de-excitation laser, 
we first apply a microwave pulse of $30~$ns to transfer atoms in $\ket{\uparrow}$ to $\ket{75D_{3/2},m_J=-3/2}$, 
which has a much smaller coupling to $\ket{\downarrow}$. This procedure leads to a freezing 
of the interaction-induced dynamics. }. 
An atom in the Rydberg state $nS_{1/2}$ is thus detected at the end of the sequence, while an atom in 
$nP_{3/2}$ state is lost. This detection technique leads to false positives with a $5\%$ probability and false 
negatives with $3.5\%$ probability~\cite{Deleseleuc2018}.  
We include the state preparation and measurement errors (SPAM) in the numerical  simulations when comparing to the data.

\section{Implementation of the XXZ Hamiltonian with two atoms}\label{Sec:2atoms}

\begin{figure}
	\includegraphics[width = 8.6 cm]{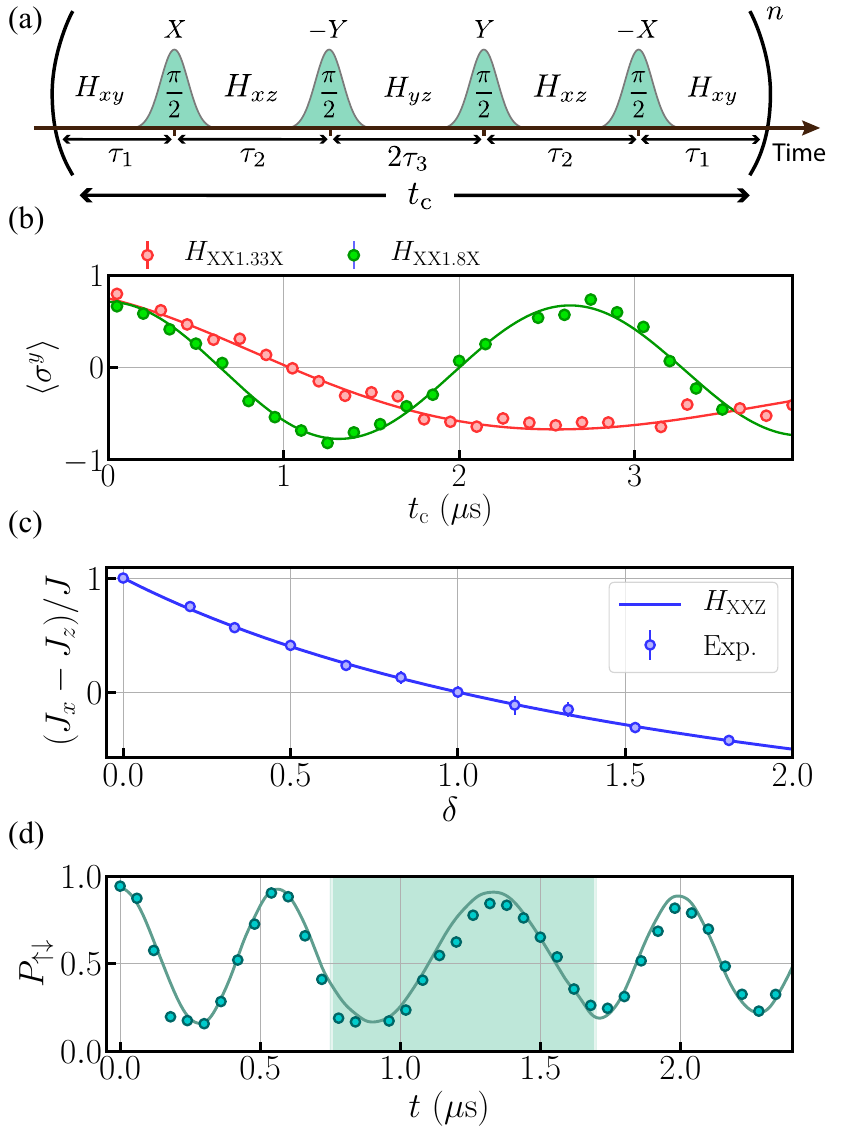}
	\caption{\textbf{Implementation of XXZ Hamiltonians with two atoms}. 
		(a) Microwave sequence consisting of four $\pi/2$-Gaussian pulses driving rotations around $ X, -Y, Y, -X$.
		(b) Evolution of the $y$-magnetization under $H_{\rm driven}$ after initialization  
		in $\ket{\rightarrow \rightarrow}_y$ as a function of $t_c$, for two ratios $\tau_1/\tau_2$ corresponding to 
		$\delta=1.33, 1.8$. The lines are fit to the data.
		(c) Normalized oscillation frequency of the $y$-magnetization as a function of $\delta$. 
		Circles: experimental results (error bars from the fits of the oscillations). 
		Solid line: prediction from Eq.~(\ref{Eq:HXXZ}) with no adjustable  parameters. 
		(d) Evolution of the probability $P_{\uparrow \downarrow}$ 
		under $H_{\rm XX}$ and $H_{\rm XXX}$ (green shaded region), following the preparation 
		in $\ket{\uparrow \downarrow}$. Solid lines: simulation using the XXZ Hamiltonian. 		
		(b,d) Error bars represent the s.e.m., often smaller than the symbol size.}
	\label{fig:twoatoms}
\end{figure}

In this section, we demonstrate the implementation of the XXZ Hamiltonian of Eq.~(\ref{Eq:HXXZ}) in 
the case of two interacting atoms. We use the pseudo-spin states
$\ket{\downarrow} = \ket{90S_{1/2}, m_J = 1/2}$ and $\ket{\uparrow} = \ket{90P_{3/2}, m_J = 3/2}$
separated by $\omega_{\rm MW}/2\pi = 5.1 \, \text{GHz}$ and coupled by the microwave field with 
a mean Rabi frequency averaged over the Gaussian pulses $\Omega = 2\pi\times7.2 \, \text{MHz}$. 
The atoms are separated by $30 \,  \mu \text{m}$, 
leading to 
$J \simeq 2\pi\times 930 \, \text{kHz}$. 

The spectrum of the XXZ Hamiltonian for two atoms consists of two 
degenerate eigenstates $\ket{\downarrow\downarrow}$ and $\ket{\uparrow\uparrow}$ with energy $J_z$
and two other eigenstates $\ket{\pm}  = (\ket{\uparrow \downarrow} \pm \ket{\downarrow \uparrow})/\sqrt{2}$
with energy $-J_z\pm 2J_x$. 
To characterize the engineering of the XXZ Hamiltonian, 
we first initialize the atoms in the state $\ket{\rightarrow \rightarrow}_y = 
(\ket{\uparrow \uparrow} - \ket{\downarrow \downarrow} + i\sqrt{2} \ket{+})/2$,
by applying a $\pi / 2$ pulse around the $x$-axis. We then apply one sequence of four microwave pulses, 
varying $t_c$ for a fixed ratio $\tau_1/\tau_2 $, {\it i.e.}, a given ani\-so\-tro\-py $\delta$.
This state evolves with time, and the  total $y$-magnetization $\langle \sigma^y \rangle$ 
oscillates at a frequency $2|J_x - J_z|$ (see Fig.~\ref{fig:twoatoms}b). 
We measure this frequency as a function of $\delta$ (see Fig.~\ref{fig:twoatoms}c) 
and find excellent agreement with the predicted value (Eq.~\ref{Eq:HXXZ}). 

To demonstrate the dynamical tunability of this microwave engineering,
we perform an experiment in which 
we change the Hamiltonian during the evolution of the system. 
We initialize the atoms in $\ket{\uparrow \downarrow}$ and measure the 
probability $P_{\uparrow\downarrow}$ as a function of time. 
We first let the system evolve under $H_{\rm XX}$ and observe an oscillation 
between  $\ket{\uparrow \downarrow}$ and $\ket{\downarrow \uparrow}$ 
at a frequency $2 J$, see Fig.~\ref{fig:twoatoms}(d). 
Between $t=0.8-1.7~\mu$s, we apply a single microwave sequence, varying $t_c$ while keeping $\tau_1 = \tau_2$
to engineer $H_{\rm XXX}$. We observe a reduction of the oscillation frequency
by a factor $0.65(2)$, in agreement with 
the expected factor of $2/3$. We then switch off the microwaves, and the exchange at frequency $2J$ resumes.
This engineering does not introduce 
extra sizable decoherence beyond the freely evolving case. 
We compare the results of the experiment with the solution of
the Schr\"odinger equation using the Hamiltonian (Eq.~\ref{Eq:HXXZ}). We  
include the residual imperfections measured on the experiment: 
SPAM and shot-to-shot fluctuations
of the interatomic distance. The results of the simulations are shown as solid lines
in Fig.~\ref{fig:twoatoms}(d), and agree well with the data. 

\section{Freezing of the magnetization in a 2D array}\label{Sec:2D}

\begin{figure}
	\includegraphics[width = 8.6 cm]{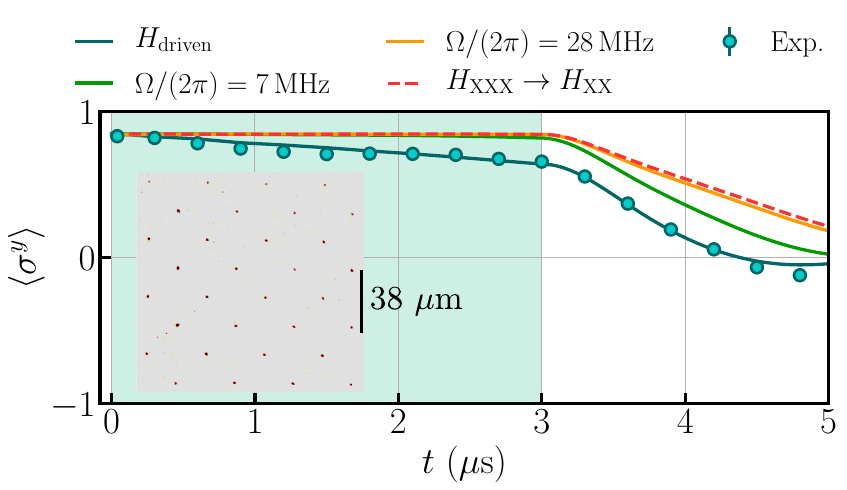}
	\caption{\textbf{Freezing of the magnetization in a 2D square array}. 
    	Evolution of the total magnetization along $y$, $\langle \sigma^y \rangle$ 
    	after initialization in $\ket{\rightarrow \rightarrow \cdots \rightarrow}_y$, and evolution under
	$H_{\rm driven}$ for the first 3~$\mu$s and $H_{\rm XX}$ afterwards. 
	Lines:  numerical simulations based on the MACE method (see text). All include SPAM errors.
	Blue: $H_{\rm driven}$ including the microwave imperfections.
	Green: without microwave imperfections.
	Orange: pulse Rabi frequency $\Omega=2\pi\times 28$~MHz, no microwave imperfection.
	Dashed red: evolution under $H_{\rm XXX}$, followed by $H_{\rm XX}$.
	Inset: fluorescence image of the 2D square array containing 32 atoms with an inter-site 
	distance $a \simeq 27 \, \mu \text{m}$, leading to a nearest-neighbor interaction 
	energy without microwaves $J \simeq 2\pi\times 133 \, \text{kHz}$ and a mean interaction 
	energy $J_m\simeq 2\pi\times 720~$kHz.}
	\label{fig:manyAtoms}
\end{figure}

We now implement the Hamiltonian engineering technique 
in a two-dimensional square array consisting of 32 atoms (see Fig.~\ref{fig:manyAtoms}). 
For this purpose, as was done in Ref.~\cite{Geier2021} for a gas of cold atoms, we engineer
the XXX Heisenberg model for which the total magnetization is a conserved quantity.
The ability to freeze the magnetization of a system for a controllable time provides a potential 
route towards dynamical decoupling and quantum sensing~\cite{Choi2020}.  

For this experiment and for those in the next section, 
we use the Rydberg states $\ket{\downarrow} = \ket{75S_{1/2}, m_J = 1/2}$ 
and $\ket{\uparrow} = \ket{75P_{3/2}, m_J = -1/2}$, separated by $\omega_{\rm MW}/2\pi=8.5$~GHz.  
We initialize the system in the $\ket{\rightarrow \rightarrow \cdots \rightarrow}_y$ state.
We apply several sequences of the driven Hamiltonian for $3 \,\mu\text{s}$ and then we switch off the drive and 
let the system evolve under $H_{\rm XX}$. We use $t_c = 300 \, \text{ns}$ and 
Gaussian microwave pulses with a $1/e^2$ width of $16.8$\,ns. 
We measure the total magnetization $\langle \sigma^y \rangle$ after the 
application of an increasing number of sequences. 
The results are shown in Fig.~\ref{fig:manyAtoms}
where, as expected, we observe an approximately constant magnetization for the first $3 \, \mu\text{s}$, followed
by its decay towards zero under $H_{\rm XX}$. This demagnetization results from the beating of all the eigenfrequencies 
of $H_{\rm XX}$ for this many-atom system.

As the ab-initio calculation of the dynamics  is now more challenging, we use a 
Moving-Average-Cluster-Expansion (MACE) method~\cite{Hazzard2014} to simulate the system. 
This method consists in diagonalizing clusters, here of 12 atoms, using the Schr\"odinger equation
and averaging the results over all 12-atom cluster configurations possible with 32 atoms. 
We include in the simulation the SPAM errors and imperfections in the 
microwave pulses calibrated on a single atom (see Appendix~\ref{App:calibMW}).
As shown in Fig.~\ref{fig:manyAtoms}, the simulation, without adjustable parameters, is in good agreement 
with the observed dynamics at all times. 
However, the comparison with the evolution under $H_{\rm XXZ}$ (red dashed line) reveals that our engineering is not perfect.

The simulation allows us to assess the contribution of various effects to explain this difference. 
First, not taking into account the imperfections of the microwave  in the simulation (green solid line) 
leads to a nearly perfect freezing of the magnetization
during the application of the pulses: the observed residual decay of the magnetization 
is thus a consequence of the microwave imperfections. 
Second, after switching off the microwave field, the dynamics under $H_{\rm XX}$ differs
depending on whether it starts from the state produced by $H_{\rm XXX}$ or $H_{\rm driven}$ at $t=3$~$\mu$s. 
This difference originates from the finite duration of the microwave pulses during which the 
interactions play a role: an average Rabi frequency four times larger than in the experiment 
($\Omega=2\pi\times 28$~MHz, orange curve), would already lead to a nearly perfect agreement between 
the evolution under  $H_{\rm XXX}$ and $H_{\rm driven}$. 
The agreement finally indicates that the value $J_m t_c\approx 2\pi\times 0.2$ is already low 
enough for a faithful implementation of the XXX model.

\section{Dynamics of domain wall states in 1D systems}\label{Sec:1Dwall}

\begin{figure*}[t!]
	\includegraphics[width = 17 cm]{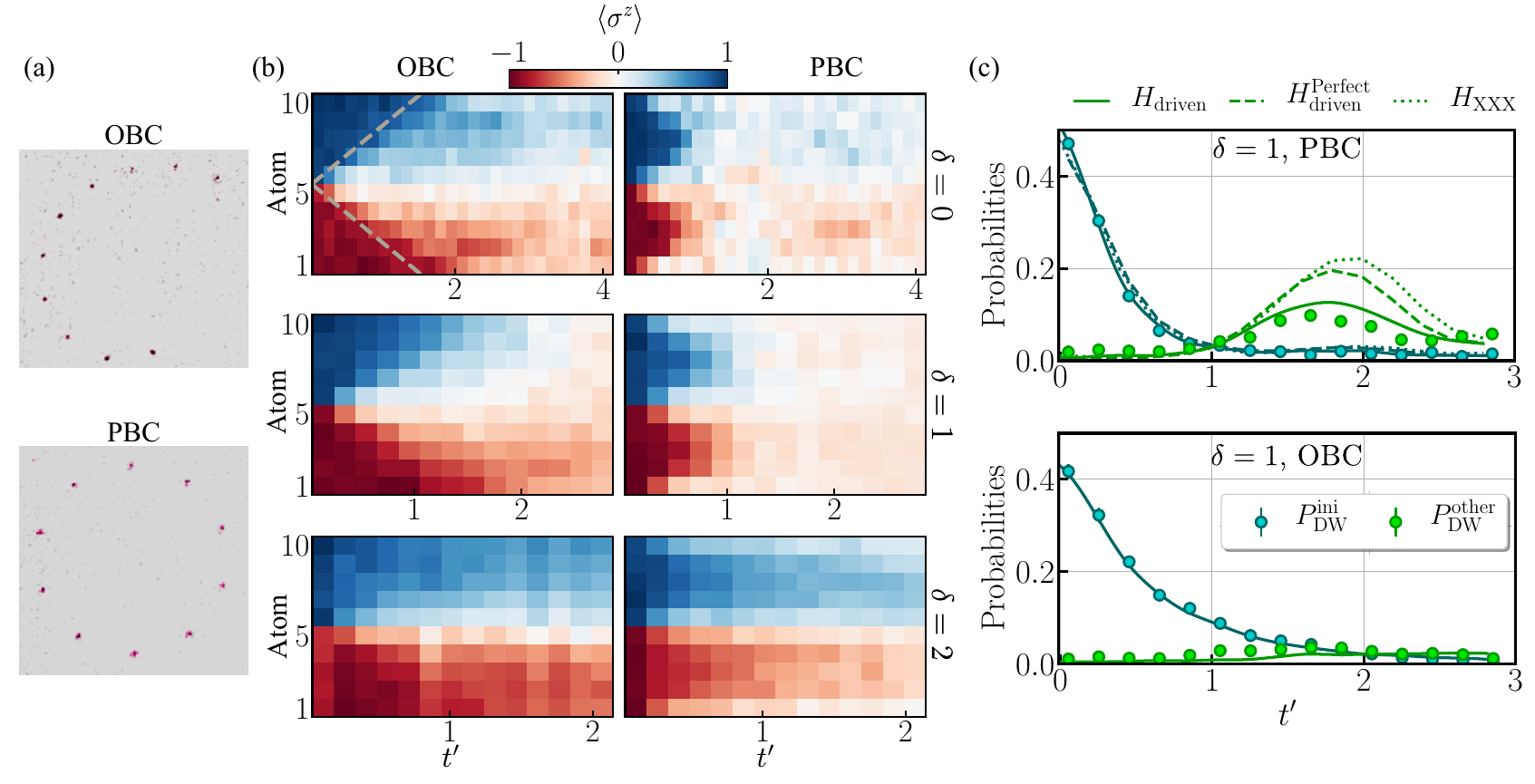}
	\caption{\textbf{Dynamics of domain wall states under $H_{\rm XXZ}$ in 1D systems}. 
	(a) Fluorescence images of the trapped atoms for the two geometries used in the experiment: 
	the spiral implements open boundary conditions (OBC), while 
	the circle realizes periodic boundary conditions (PBC). 
	(b) Density maps of the temporal evolution of the $z$-magnetization $\langle\sigma_i^z\rangle$ as a 
	function of the normalized time $t'$, 
	following the preparation of a 
	domain wall state, for anisotropies $\delta=0,1,2$, and  OBC (left) and PBC (right) geometries. 
	The dashed grey line shows the light-cone $\xi= \pm 2 J t^\prime$. 
	(c) Evolution of the probability of occurence of domain walls for OBC and PBC. 
	The solid lines are the simulations using $H_{\rm driven}$ accounting for all experimental imperfections
	(SPAM errors, shot-to-shot fluctuations in atomic positions and microwave imperfections). 
	The dashed (dotted) line is a simulation using $H_{\rm driven}^{\rm Perfect}$ ($H_{\rm XXZ}$) without 
	microwave imperfections.}
	\label{fig:domainWall}
\end{figure*}

In a last set of experiments, we illustrate the engineering of $H_{\rm XXZ}$ Hamiltonians 
on the dynamics of a domain wall (DW), {\it i.e.}, a situation where a boundary 
separates spin-up atoms from spin-down ones,  in a one-dimensional chain with periodic (PBC) or open (OBC) 
boundary conditions. Transport properties in the nearest-neighbor XXZ model and for large system sizes 
have been studied extensively, both analytically and numerically. The evolution of such a system depends on $\delta$ 
due to two competing effects: a melting of the DW caused by spin-flips with a rate $J_x$, and an opposing associated 
energy cost of $2J_z$, which maintains the DW. In the case of a pure initial state 
(the relevant situation for our experiment), for $\delta<1$, 
the domain-wall is predicted to melt, with a magnetization profile expanding ballistically in 
time\,\cite{Collura18,Misguich2019}. At the isotropic point ($\delta = 1$), one expects a diffusive behavior with 
logarithmic corrections\,\cite{Misguich2017}. For $\delta>1$, the magnetization profile 
should be frozen at long times \cite{Gobert2005,Mossel2010,Misguich2019}. 
All these theoretical predictions have been explored for large system sizes. 

Here, we study the emergence of these properties with a few-body system of 10 atoms with 
interatomic distance $a = 19 \, \mu \text{m}$ (see Fig.~\ref{fig:domainWall}a). This yields a nearest-neighbor interaction 
$J \simeq 2\pi\times 270 \, \text{kHz}$ and $J_m\simeq 2\pi\times 0.6~$MHz, 
which fulfills the condition  $J_m t_c\ll 2\pi$ for $t_c=300$~ns. 
Using the addressing technique described in Sec.\ref{Sec:Expsetup}, we prepare 
five adjacent atoms in $\ket{\uparrow}$ and the remaining ones in $\ket{\downarrow}$.
We then study the evolution of the system under $H_{\rm XXZ}$ for different $\delta$.

We first look at the evolution of the single-site magnetization $\langle \sigma_i^z \rangle$ as a function of 
the normalized time $t'=t\,J_x(\delta) /(J\times 1\mu \rm{s})$.
The results for OBC are shown in Fig.\ref{fig:domainWall}(b) 
with $\delta = 0, 1, 2$ \footnote{Implementing $H_{\rm XX2X}$ requires $\tau_1=0$. 
We therefore remove the $X$ and $-X$ pulses from the sequence, with the exception of the first and final pulses. }. 
For $\delta \leq 1$, we  observe the melting of the domain wall, 
resulting in an approximately uniform magnetization profile for $t'\gtrsim 3$. 
In the case $\delta = 0$, the width $2\xi$ of the magnetization profile grows 
ballistically in time, as predicted, and  follows a light-cone dynamics, $\xi= \pm 2 J \,t'$~\cite{Collura18,Misguich2019}, 
illustrated by the dashed grey lines in the top left panel of Fig.~\ref{fig:domainWall}(b). 
At the isotropic point $\delta = 1$, the melting of the wall happens more slowly, 
as the cost of breaking the spin domains becomes higher. For $\delta = 2$, 
we observe a retention of the domain wall at all times: 
the magnetization profile hardly  evolves between $t^\prime=1.1$ and $t^\prime=2.0$, 
indicating a freezing of the system dynamics. Our Hamiltonian engineering is thus able 
to distinguish different spin-transport behaviors for various $\delta$. 

We now consider the case where the atoms are arranged in a circle (PBC). 
One expects comparable behavior as for the OBC case, with 
the two domain walls melting ballistically for $\delta=0$, and more slowly for increasing $\delta$.
This is what we observe in Fig.~\ref{fig:domainWall}(b) 
with the system reaching a depolarized state more quickly 
than for the OBC due to the presence of two edges. 
However, the  dynamics for $t'\gtrsim 1$ differs between PBC and OBC when considering as
an observable the probability $P_{\rm DW}$ to observe a given domain wall, as we now illustrate for
the case $\delta=1$. The probability $P_{\rm DW}$ is defined as the probability to find a cluster of adjacent $\ket{\uparrow}$ 
excitations in the chain after an evolution time.
The results for the two boundary conditions are shown in Fig.~\ref{fig:domainWall}(c) where we 
plot the probability $P_{\rm DW}^{\rm ini}$
to find the {\it initial} domain wall after an evolution time $t'$
\footnote{To increase the amplitude of the revival in $P_{\rm DW}^{\rm other}$ observed in Fig.~\ref{fig:domainWall}(c)
we include in the data and in the simulations events containing domain walls with 4, 5 and 6 excitations.}. 
We do observe the melting of the initial wall, and the fact that it disappears faster for PBC than for OBC.   
We also plot the probability $P_{\rm DW}^{\rm other}$ to find a domain wall at a location {\it different} from the initial one. 
Interestingly, for PBC, while the average magnetization has reached equilibrium (Fig.~\ref{fig:domainWall}b) and the initial  wall 
has melted, $P_{\rm DW}^{\rm other}$ still evolves: 
domain walls appear at different locations around the circle
for $t'\approx1.7$
(see also simulations for longer times in App. \ref{App:XXX}). 
The OBC case shows a much weaker transfer of the initial domain wall towards other ones, thus
revealing the role of the boundary conditions (see Fig. \ref{FigApp:XXX}).

\begin{figure}
	\includegraphics[width = 8.6 cm]{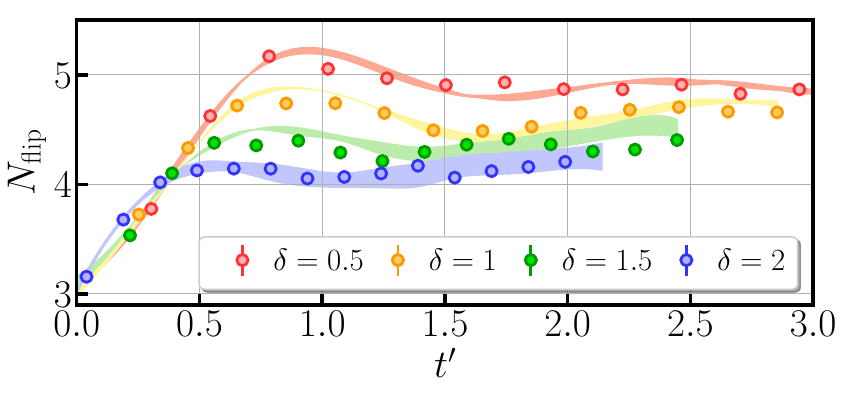}
	\caption{\textbf{Dynamics of the number of spin flips (PBC)}. 
	Evolution of $N_{\rm flip}$ for $\delta=0.5, 1, 1.5, 2$ as a function 
	of the normalized time $t'$. 
	Shaded regions: results of the simulation using $H_{\rm driven}$, including the $6\pm1\%$ fluctuations 
	on the microwave rotation axis (see Appendix~\ref{App:calibMW}). }
	\label{fig:delocalization}
\end{figure}

To further understand the domain wall structure around the circle (PBC), we consider 
the  spin correlations $\langle \sigma_i^z \sigma_{i+1}^z \rangle$, related
to the number of spin-flips $N_{\text{flip}}$ by:
\begin{equation}
N_{\text{flip}} = {1\over 2}\sum_i 1-\langle \sigma_i^z \sigma_{i+1}^z \rangle
\end{equation} 
(a flip is defined as two neighboring atoms in opposite spin states).
The initialized DW state would therefore consist of two spin flips, while a fully 
uncorrelated state contains $N/2$ on average.  
We show in Fig.~\ref{fig:delocalization} the dynamics of $N_{\text{flip}} $
for four values of the anisotropy for PBC. 
For $\delta <  1$, $N_{\rm flip}$ approaches $N/2$ at long time, confirming the fact that the 
system becomes fully uncorrelated. 
However, for increasing $\delta$, the value of $N_{\rm flip}$ at long times decreases. This means 
that the $\ket{\uparrow}$ excitations tend to remain bunched for large $\delta$.

We finally compare the experimental data shown in Figs.~\ref{fig:domainWall}(c) and \ref{fig:delocalization}
with numerical simulations using both $H_{\rm driven}$ and 
the target $H_{\rm XXZ}$ Hamiltonian. For both we include SPAM errors, which are chosen to match the initial state, 
the residual shot-to-shot fluctuations of the interatomic distances, and the microwave imperfections for $H_{\rm driven}$.
The results for the simulation of the probability of domain wall are shown in Fig.~\ref{fig:domainWall}(c):
the data are well approximated by the $H_{\rm driven}$ simulation, 
indicating that we understand the sources of experimental errors. However, not including the microwave pulses 
imperfections in the simulation ($H_{\rm driven}^{\rm Perfect}$ in Fig.~\ref{fig:domainWall}c)  reveals a difference
with the dynamics driven by $H_{\rm XXZ}$. We show in the Appendix~\ref{App:XXX} that this originates from the
finite duration of the pulses during which the interaction play a role, as observed in Sec.\ref{Sec:2D}. 
We also plot in Fig.~\ref{fig:delocalization} the simulation of $N_{\rm flip}$ using $H_{\rm driven}$, 
including all the imperfections and find good agreement with the data.

\section{Conclusion}

In this work, we have engineered  XXZ Hamiltonians with anisotropies $0\leq\delta\leq2$
using the resonant dipole-dipole interaction between Rydberg atoms in arrays 
coupled to a resonant microwave  field. We have illustrated the method on two iconic situations: 
the Heisenberg model in 2D square arrays, where we demonstrate the ability to dynamically freeze the evolution of
a state with a given magnetization, and the dynamics  of a domain wall in a 1D chain 
with open and periodic boundary conditions.   
By comparing our results to numerical simulations we infer the two
current limitations on our setup: 
(i) the imperfections in the 8.5 GHz microwave pulses, and
(ii) the lack of microwave power that prevents us from reaching pulses 
short enough to be able to neglect the residual influence of the interactions during their application.
Despite these limitations, which can be solved by improving the microwave hardware, we were able to observe 
all the qualitative features of the situations we explored. This highlights the versatility of a Rydberg-based quantum simulator, 
beyond the implementation of the natural Ising-like or XX Hamiltonians.
Future work could include the study of frustration in various arrays governed by the Heisenberg model~\cite{Richter2004},
or the study of domain wall dynamics for larger system size to confirm the various delocalization 
scalings beyond the emergent behaviors studied here. We also anticipate that  combining  
microwave drive with the ability to locally address the resonance frequency of the atoms using light-shifts 
would lead to the engineering of a broader class  of Hamiltonians. 

\begin{acknowledgments}
This work is supported by the European Union's Horizon 2020 research and innovation program 
under grant agreement no. 817482 (PASQuanS), the Agence National de la Recherche (ANR, project RYBOTIN),
the Deutsche Forschungsgemeinschaft (DFG, German Research Foundation) 
under Germany's Excellence Strategy EXC2181/1-390900948 (the Heidelberg STRUCTURES 
Excellence Cluster), within the Collaborative Research Center SFB1225 (ISOQUANT), 
the DFG Priority Program 1929 ``GiRyd" (DFG WE2661/12-1), and by the Heidelberg Center for Quantum Dynamics. 
C.H. acknowledges funding from the Alexander von Humboldt foundation, T.F. from a graduate 
scholarship of the Heidelberg University (LGFG), and 
D.B. from the Ram\' on y Cajal program (RYC2018-025348-I). 
F.W. is partially supported by the Erasmus+ program of the EU.
The authors also acknowledge support by the state of 
Baden-W\"urttemberg through bwHPC and the German Research Foundation (DFG) 
through grant no INST 40/575-1 FUGG (JUSTUS 2 cluster).
\end{acknowledgments}
\clearpage

\setcounter{figure}{0}
\renewcommand\thefigure{A\arabic{figure}} 
\setcounter{equation}{0}
\renewcommand\theequation{A\arabic{equation}} 
\appendix

\section{Calibration of the microwave pulse sequence on a single atom}\label{App:calibMW}

\begin{figure}
	\includegraphics[width=8.6 cm]{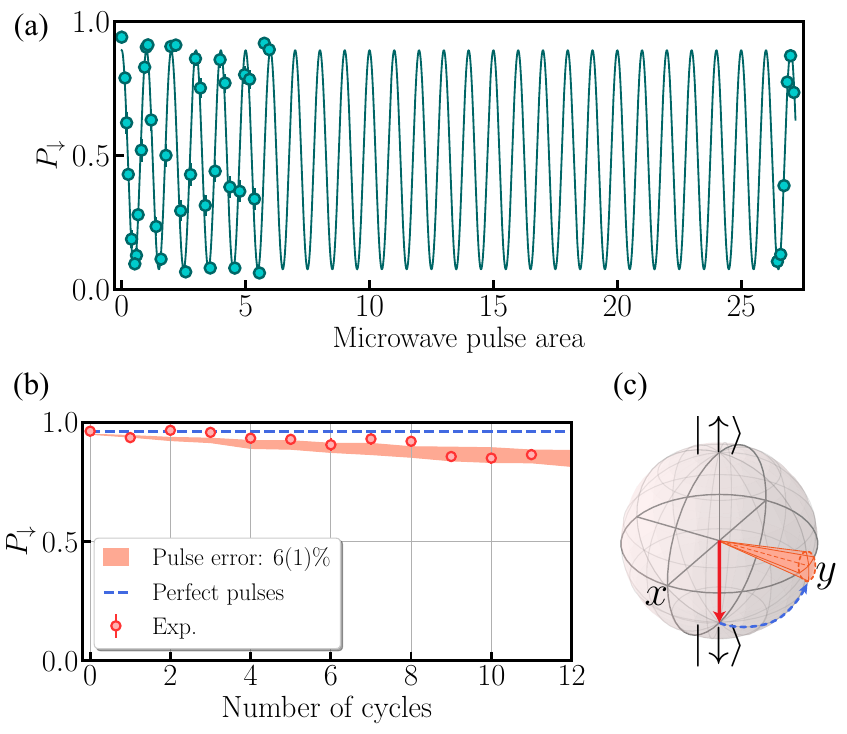}
	\caption{\textbf{Calibration of microwave pulse error.} 
		(a) Microwave Rabi oscillations between states $\ket{\uparrow}$ and $\ket{\downarrow}$. The Rabi frequency
		is $\Omega=2\pi \times 13.2$~MHz.
		(b) Probability $P_{\downarrow}$ of measuring a single atom 
		in $\ket{\downarrow}$ following $H_{\rm XXX}$ versus the number of cycles. 
		The data is shown as red circles, with the simulated pulse sequence results 
		for perfect pulses shown in blue and with a pulse error $\Delta\theta=0.06\pm0.01$ 
		as the shaded pink region. 
		(c) Illustration of how the error is included in the simulation: the final Bloch
		vector after a rotation around any axis (here the $x$-axis)
		lies inside the orange cone.  }
	\label{fig:single_atom}
\end{figure}

The microwave field is sent onto the atoms using a microwave antenna, 
with poor control over the polarization due to the presence of metallic parts surrounding the atoms. 
An example of Rabi oscillation on the $\ket{\downarrow}-\ket{\uparrow}$ transition using a long microwave pulse
is shown in Fig.~\ref{fig:single_atom}(a). We observe no appreciable damping after 25 oscillations. 

To implement $H_{\rm driven}$, we have empirically found that applying pulses with Gaussian, 
rather than square envelopes minimises pulse errors arising from the fast switch on/off.  
In order to assess the influence of further imperfections in the microwave pulses on the dynamics of 
the systems used in this work, we compare {\it single} atom data with a numerical simulation. 
We prepare an atom in $\ket{\downarrow}=\ket{75S_{1/2}, m_J = 1/2}$ and then implement sequences of four 
$\pi/2$- Gaussian pulses, in the same way as for the many-body system. Following a single, four-pulse cycle 
one would expect the atom to have returned to  $\ket{\downarrow}$. Figure \ref{fig:single_atom}(b) 
shows the probability of measuring the atom in $\ket{\downarrow}$ after each cycle, where we 
see a slow decrease in $P_{\downarrow}$. 

In the main text, we concluded  that part of the discrepancy between the experimental 
results and the prediction of the XXZ Hamiltonian simulation 
came from errors in the microwave pulses. The source of these errors could be fluctuations in 
the amplitude and/or the phase of the microwave pulses, 
difficult to measure at frequencies in the 5-10 GHz range. 
To encompass these effects, we phenomenologically include in our simulations an uncertainty in the 
angle of rotation of the microwave pulse: for each pulse, we assign two values 
$n_1$ and $n_2$ from a normal distribution centered around zero with a standard deviation $\Delta\theta$. 
We then use these values to describe the rotation operator: if the desired rotation axis is 
$x$, the actual rotation is performed around the axis $x^\prime$ such that
\begin{equation}
\sigma^{x^\prime} = (1-n_1^2-n_2^2)^{1/2}\sigma^x + n_1\sigma^y + n_2\sigma^z.
\end{equation}
This effect is illustrated in Fig.~\ref{fig:single_atom}(c). In Fig.~\ref{fig:single_atom}(b) 
the shaded area shows an uncertainty in pulse error of $\Delta\theta = 0.06\pm0.01$, which closely 
matches the experimental results. We use this value of the  uncertainty in all the many-body simulations using 
$H_{\rm driven}$ presented in this work. 

\section{Influence of the finite duration of the microwave pulses on the simulated many-body Hamiltonian}\label{App:XXX}

\begin{figure}
	\includegraphics[width=8.6cm]{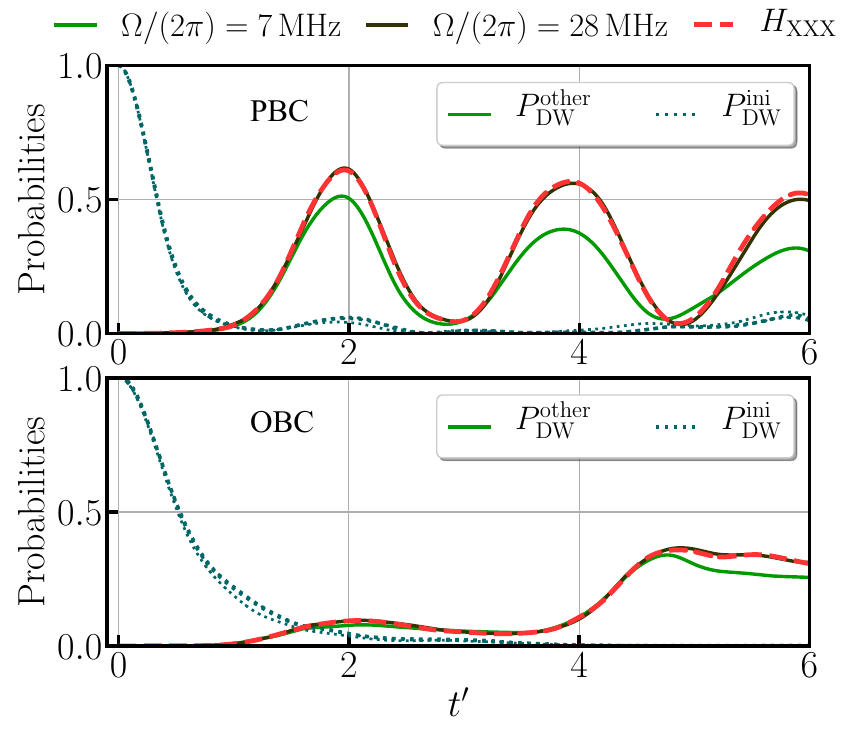}
	\caption{\textbf{Influence of the finite duration of the microwave pulses.}
	Comparison between the evolution of the probabilities of domain walls
	under $H_{\rm driven}$ and $H_{\rm XXX}$ in PBC (upper) and OBC (lower).
	The simulations do not include any experimental error to highlight the role of the finite 
	duration of the microwave pulses, with average Rabi frequency $\Omega$. 
	The large increase of $P_{\rm DW}$ observed for $t'>4$ in OBC is a consequence of the reflexion of the excitations 
	at the edges of the chain.}
	\label{FigApp:XXX}
\end{figure}

We have seen in Sec.\ref{Sec:2D} that, in the case of the 2D array, increasing the Rabi frequency 
of the microwave pulses by a factor four, {\it i.e.}, decreasing their duration by four, 
leads to a nearly perfect agreement  between the evolution under $H_{\rm driven}$ and $H_{\rm XXX}$. 
We perform here the same analysis for the evolution of the domain wall for both the 
periodic and open boundary conditions (Sec.\ref{Sec:1Dwall}). 
The results of the simulation of the probability of domain wall
{\it without} experimental imperfections is shown in Fig.~\ref{FigApp:XXX} for an 
evolution longer than the one achieved in the experiment and for two Rabi frequencies. Similarly to the 2D case, 
a Rabi frequency four times larger than the one we could reach in the experiment would lead to a very good agreement 
with the evolution under $H_{\rm XXX}$. This simulation indicates here also that the value $J_m t_c \approx 2\pi \times 0.2$
is low enough and that the use of discrete pulses does not thwart a faithful implementation of the XXZ Hamiltonian.

%
\end{document}